\documentclass[12pt,preprint]{aastex}
\newcommand{\cluster}{RCS043938-2904.9} 
\newcommand{\clusterz}{$z=0.951$}
\newcommand{\etal}{\it et al.}  
\newcommand{\I}{$I$}
\newcommand{\Js}{$J_s$}
\newcommand{\Ks}{$K_s$} 
\slugcomment{20-10-2004}
\begin{document}

\title{RCS043938-2904.9: A New Rich Cluster of Galaxies at 
$z=0.951$\altaffilmark{1}}

\author{
L. Felipe Barrientos\altaffilmark{2},
Michael D. Gladders\altaffilmark{3},
H.K.C. Yee\altaffilmark{4},
Leopoldo Infante\altaffilmark{2},
Erica Ellingson\altaffilmark{6}, 
Patrick B. Hall\altaffilmark{2,5} and
Gisela Hertling\altaffilmark{2}
}
\altaffiltext{1}{Based on observations obtained at: (a) the Cerro Tololo
Inter-American Observatory.  CTIO is operated by AURA, Inc.\ under
contract to the National Science Foundation; (b) The European Southern
Observatory, Paranal, Chile; and (c) the Observatories of the Carnegie
Institution of Washington, Las Campanas, Chile.}

\altaffiltext{2}{Departamento de Astronom\'{\i}a y Astrof\'{\i}sica,
Pontificia Universidad Cat\'olica de Chile, Avda. Vicu\~na Mackenna
4860, Casilla 306, Santiago 22, Chile; barrientos@astro.puc.cl,
phall@astro.puc.cl, linfante@astro.puc.cl}

\altaffiltext{3}{Observatories of the Carnegie Institution of
Washington, 813 Santa Barbara Street, Pasadena, California 91101, USA;
gladders@ociw.edu} 

\altaffiltext{4}{Department of Astronomy and Astrophysics, University
of Toronto, 60 St. George Street, Toronto, ON M5S 3H8, Canada;
hyee@astro.utoronto.ca}

\altaffiltext{5}{Princeton University Observatory, Peyton Hall,
Princeton, NJ 08544, USA}

\altaffiltext{6}{Center for Astrophysics and Space Astronomy, CB389
University of Colorado, Boulder, CO 80309, USA;
e.elling@casa.colorado.edu}

\begin{abstract}
We present deep $I,J_s,K_s$ imaging and optical spectroscopy of the
newly discovered Red-Sequence Cluster Survey cluster \cluster. This
cluster, drawn from an extensive preliminary list, was selected for
detailed study on the basis of its apparent optical
richness. Spectroscopy of 11 members places the cluster at \clusterz
$\pm 0.006$, and confirms the photometric redshift estimate from the
(R-z) color-magnitude diagram. Analysis of the infrared imaging data
demonstrates that the cluster is extremely rich, with excess counts in
the \Ks-band exceeding the expected background counts by
$9\sigma$. The properties of the galaxies in \cluster\ are consistent
with those seen in other clusters at similar redshifts. Specifically,
the red-sequence color, slope and scatter, and the size-magnitude
relation of these galaxies are all consistent with that seen in the
few other high redshift clusters known, and indeed are consistent with
appropriately evolved properties of local cluster galaxies. The
apparent consistency of these systems implies that the rich,
high-redshift RCS clusters are directly comparable to the few other
systems known at $z \sim 1$, most of which have been selected on the
basis of X-ray emission.

\end{abstract}


\section{Introduction}
The search for distant galaxy clusters began decades ago with
\cite{abe58} original identification of a large sample of (then)
distant clusters in the Palomar Optical Sky Survey plates. A major
goal of many surveys since has been to push the redshift limits of
searches to as high as possible. Currently, the most distant clusters
known are at $z \sim 1.3$ \citep{stanford1997,rosati1999} . A wide
variety of techniques, both blank field \citep[e.g.,][]{dalcanton1996}
and targeted \citep[e.g.,][]{chambers1996} have been used as viable
methods for pushing cluster searches to high redshifts, with the
blank-field methods being the most promising for cosmologically
motivated studies. Broadly, these methods either search for visible
effects from the intra-cluster gas via X-ray surveys
\citep[e.g.,][]{rosati1999,mul03} or proposed Sunyaev-Zeldovich (SZ)
effect surveys, or the effects of the cluster galaxies themselves, via
extragalactic background light fluctuations \citep{dalcanton1996}, or
matched-filter \citep[e.g.,][]{pos96} or cluster red-sequence
\citep{gladders2000} analyses of resolved images. Direct detection of
massive clusters via weak-lensing has also been suggested
\citep{wittman2001}.

A significant, and at this point unresolved, issue is whether or not
the clusters discovered by various methods will in fact sample the
same fundamental population. This is particularly relevant at high
redshifts since various observables are likely to be more divergent
and less closely tied to mass during the process of cluster
assembly. Addressing these issues fully requires surveys using
multiple techniques over the same areas. In this letter we present
detailed observations of one of the first and richest clusters found
in the RCS at $z \sim 1$. We use $\Omega_M=0.27$, $\Omega_\Lambda =
0.73$ and $H_0=71$ km~s$^{-1}$~Mpc$^{-1}$ throughout.

\section{The Red-Sequence Cluster Survey}

The Red-Sequence Cluster Survey (RCS) searches for clusters by
exploring over-densities in position and color space, detecting the
unambiguous signal from the early-type galaxy color sequence seen in
local and high redshift clusters \citep{gladders2000}. The survey has
been conducted using the large format mosaic cameras at CFHT and the
CTIO 4m in $R(AB)$ and $z'(AB)$ to a $5~\sigma$ point source detection
limit of 25.1 and 23.7 mag, respectively. The first data and cluster
catalogs for the northern patches are presented in \citet{gladders2004}
and for the southern patches in \citet{barrientos2004}.

This letter presents detailed optical and IR observations of \cluster\
at $\alpha = 04^h39^m38\fs0, ~~\delta = -29\arcdeg 04' 55\farcs2$
(J2000), one of the most massive and distant RCS clusters for which
ancillary data have been acquired. We have detected this cluster with
a peak significance of 4.98 $\sigma$, and estimated its photometric
redshift to be $z_{phot} = 1.037$.

\section{Observations and Basic Reductions}
\subsection{Imaging}

IR imaging was taken using ISAAC on the VLT during the nights of the
26-27 October 2001, under photometric and excellent seeing conditions
($\leqslant 0\farcs5$).The exposure time was 3240 s and 5184 s in the
\Js\ and \Ks\ bands, respectively. The reduction of the images was
carried out in the usual manner, masking out the objects to properly
subtract the sky in the images. The final individual registered images
were scaled to correct for airmass differences and then combined to
produce the final \Js\ and \Ks\ images. The images were calibrated
using the IR standard stars of \citet{persson1998} to a 0.04 mag
uncertainty in \Js\ and 0.05 mag in \Ks. The $I$-band image was taken
at the Baade 6.5m telescope at the Las Campanas Observatory the night
of the 11th of January of 2002, under photometric conditions and 0.6
arcsec seeing. The $I$-band data are calibrated to \cite{lan92}
standards, with an uncertainty of 0.04 mag. The total exposure time
was 2400 s. A color composite image using the optical and IR imaging
of this cluster is shown in Figure 1.

Photometry on the field was performed using SExtractor
\citep{bertin1996} v2.2.1, using the \Ks-band image for object
detection. The total magnitude is determined using the {\em BEST}
magnitude given by Sextractor while colors are measured in a fixed
$1\farcs5$ circular diameter aperture. The IR images were convolved
with a gaussian kernel to match the seeing of the optical image in
order to determine accurate colors. In the $2.5 \times 2.5$ arcmin
area imaged in $K_s$ we find 228 galaxies down to \Ks(AB)=23, while
for a blank field we expect to find only 126 \citep{totani2001}. In
\Ks\ the cluster is an over-density of $9\sigma$, assuming Poisson
statistics for the background counts. Using the $z'$-band data from
the survey itself, we measure a red-sequence richness parameter
$B_{gcR}$ \citep[see][]{gladders2004} of $1165 \pm 399$ Mpc$^{-1.8}$
$h_{50}^{1.8}$, indicating a richness between Abell Richness 1 and 2
\citep[see][]{yee2003}, comparable to the Coma cluster.

\subsection{Spectroscopy}

Multi-object spectroscopy was carried out first on the Baade 6.5 m
telescope using LDSS2 under exceptional seeing conditions. These
observations confirmed several members. Further spectroscopy was
carried out on the VLT in service mode the nights of the 1st, 8th and
29th of January 2003. The spectra were taken using FORS2 with the
GRIS\_300I+21 grism and the OG590 filter, with a total integration
time of 9510 s (split into six observations). Galaxies were selected
by their colors in the central regions where deep IR imaging was
available, and by having similar magnitudes in the outer
regions.

One aperture mask was made for the field consisting of 37 slits with
variable lengths, typically $7 \arcsec$ and widths of $0.7 \arcsec$.
The spectra were reduced with IRAF in the usual manner.  For typical
spectra the wavelength range spans from $\sim6500$ \AA\ to $\sim9000$
\AA\, with a dispersion of $3.2$ \AA\ pix$^{-1}$, and an rms
uncertainty in the wavelength solution of 0.12 \AA.
 
Redshifts were determined using two methods. One was a line by line
absorption and emission identification and the other via
cross-correlation. For the template spectrum we used NGC 1426, NGC
1407 and a synthetic spectrum \citep{quintana1996}. Prior to
cross-correlation, all real emission and residual night sky emission
line features were removed through interpolation, and a high order
polynomial was fit to the spectra for continuum removal. A simple
parabola was fit to the correlation peak. For spectra whose
\citet{tonry1979} {\em R} value was below 3.0, a comparison was made
with absorption line identifications. If no absorption lines could be
identified in the spectra to confirm the RV result, the velocity was
discarded.

A total of 29 spectra yielded reliable redshifts, of which 11 are
consistent with being at the cluster redshift. We find that \cluster\
is at \clusterz $\pm 0.006$, while we had estimated $z=1.037$ from the
photometry alone. The spectra for one of the brightest central
galaxies is shown in Figure 2. Further details on the spectroscopy for
this and other clusters at similar redshifts will be presented in
\cite{barrientos2004}.
 
\section{Photometry Analysis}

\subsection{Size-Magnitude Relation}

The high quality VLT images enable morphological studies of the
galaxies in the field of \cluster. For this purpose we used a 2-D
galaxy light profile fitting technique \citep{schade1997a}. The
morphology and effective radius, for E/S0s, is determined by
inspecting the 2-D residual images from subtraction of a PSF-convolved
model. The model can be a bulge (i.e., an r$^{1/4}$-law), an
exponential disk, or a linear combination of both. Figure 3 shows
$r_e$ vs. $M_B(AB)$ for the morphologically selected E/S0s in the
field of \cluster. These galaxies follow a size-magnitude relation
similar to that for local cluster E/S0s. The difference between these
two populations is best interpreted as an offset in magnitude for the
galaxies in \cluster, explained as the expected luminosity change for
these galaxies due to passive evolution.

\subsection{Color-Magnitude Diagram}

Figure 4 shows the IR and IR-optical color magnitude diagrams for the
objects in the field of \cluster, within a diameter of 500 proper
kpc. This diagram shows an excess of galaxies at $(J-K)(AB) \sim 0.9$
and $(I-K)(AB) \sim 2.5$ over a wide range of magnitude. This excess
is enhanced when only the morphologically selected E/S0 galaxies (see
\S4 above) are included (solid squares). The broken line shows the
color-magnitude relation for the E/S0 galaxies in Coma, $k$-corrected
to \clusterz. The galaxies in \cluster\ are bluer than the galaxies in
Coma by $\sim 0.1$ mag in $(J-K)$ and about 0.25 in $(I-K)$. A more
detailed analysis using population synthesis models
\citep{bruzual1993} suggest we should expects a color difference of
$\Delta (I-K)=0.25$ and $\Delta (I-J)=0.09$ if most of the stars in
these galaxies were formed by $z=2$, and to $\Delta (I-K)=0.15$ and
$\Delta (I-J)=0.06$ if these were formed by $z=3$, assuming solar
metallicity. These data are consistent with a simple stellar
population formed at $z > 2$, consistent with studies of other
clusters at similar redshifts.

We use an iterated linear fit with 3-$\sigma$ clipping to fit the
red-sequence and estimate the slope and color scatter for all galaxies
with an elliptical morphology in Figure 4, in a manner akin to
\cite{gladders1998}. Errors were estimated using bootstrap re-sampling
of the input catalogs. The measured slopes are
-0.052$^{+0.025}_{-0.025}$ and -0.038$^{+0.053}_{-0.039}$ in $J-K$ and
$I-K$ respectively, and the measured scatters in the color-magnitude
relations for the E/S0 galaxies in \cluster\ amount to $\delta (J-K) =
0.056^{+0.012}_{-0.014}$ mag and $\delta (I-K) =
0.108^{+0.031}_{-0.029}$. Fitting using other techniques such as the
biweight estimator of \cite{bee90} yields similar values.  Neither
method makes an explicit background correction, nor have we subtracted
the effects of photometric errors, and so formally the reported
scatter in the red-sequence should be treated as an upper limit.

\section{Comparison to Other Clusters}

The RCS search technique relies on the presence of a population of
E/S0 galaxies with similar colors, and as such it is not surprising to
find a well defined color-magnitude sequence for \cluster. This
relation is also found in the optical/IR selected z=1.273 cluster ClG
J0848+4453 \citep{vandokkum2001} with a scatter and slope similar to
that found in \cluster. Studies of the X-ray selected clusters RDCS
J0910+5422 at z=1.106 \citep{stanford2002} and RDCS J1252.9-2927 at
z=1.237\citep{lid03} also show a similar color-magnitude relation,
with a small intrinsic scatter. The red-sequence properties seen in
\cluster\ are also consistent with those seen in the heterogeneous and
mostly lower-redshift cluster sample of \cite{stanford1998}.

Similarly, the luminosity evolution observed in \cluster\ has been
observed in other cluster samples up to $z=1.3$. The brightening of
$\Delta M_B = -1.2 \pm 0.1$ found here is consistent with that found
by a number of other studies of both the fundamental plane
\citep[e.g.,][and references therein]{vandok03} and its projection as
the Kormendy relation \citep[e.g.,][]{schade1997b}. From both the
red-sequence color, slope and scatter, and the apparent luminosity
evolution of these galaxies, we thus conclude that there is no
evidence that the red galaxy population in \cluster\ is any different
than that seen in clusters at similar redshifts but found by other
means.

\section{Conclusions}
\Ks\ imaging of \cluster\ shows and excess of 102 galaxies
($\sim9\sigma$) over the expected background counts within the imaging
field, and confirms expectations from our initial RCS survey imaging
that \cluster\ is extremely rich. Spectroscopy of \cluster\ shows it
to be a true cluster at \clusterz $\pm 0.006$, making it one of the
more distant rich cluster systems known to date. A detailed analysis
of our imaging data shows that \cluster\ has an early-type galaxy
population consistent with that seen in the few other clusters studied
at a similar redshift, in that the color-magnitude and size-magnitude
relations of these galaxies is as for other $z\sim1$ systems. This is
despite the fact that these various clusters were selected by rather
different means. Though this result is expected since the RCS is
thought to be complete for massive systems at $z=1$, the apparent
uniformity of the galaxy populations between clusters selected at
different means implies that the RCS selection method does not
introduce significant biases. This will be confirmed by on-going and
future follow-up observations in both multi-wavelength imaging and
spectroscopy of a much larger sample of $z=1$ RCS clusters.

\acknowledgments We thank the staff at LCO, CTIO and VLT for the
excellent support in the observations. LFB's research is supported by
FONDECYT, Chile, under proyecto \#1040423. This research is also
supported partially by Proyecto FONDAP ``Center for Astrophysics''
\#15010003.

\clearpage

\begin{figure}
\caption{\I\Js\Ks\ color composite image of the field centered on \cluster
($2.4 \times 2.1$ arcmin). North is up and East to the left. This
image shows approximately the central 1 Mpc.}
\end{figure}

\begin{figure}
\plotone{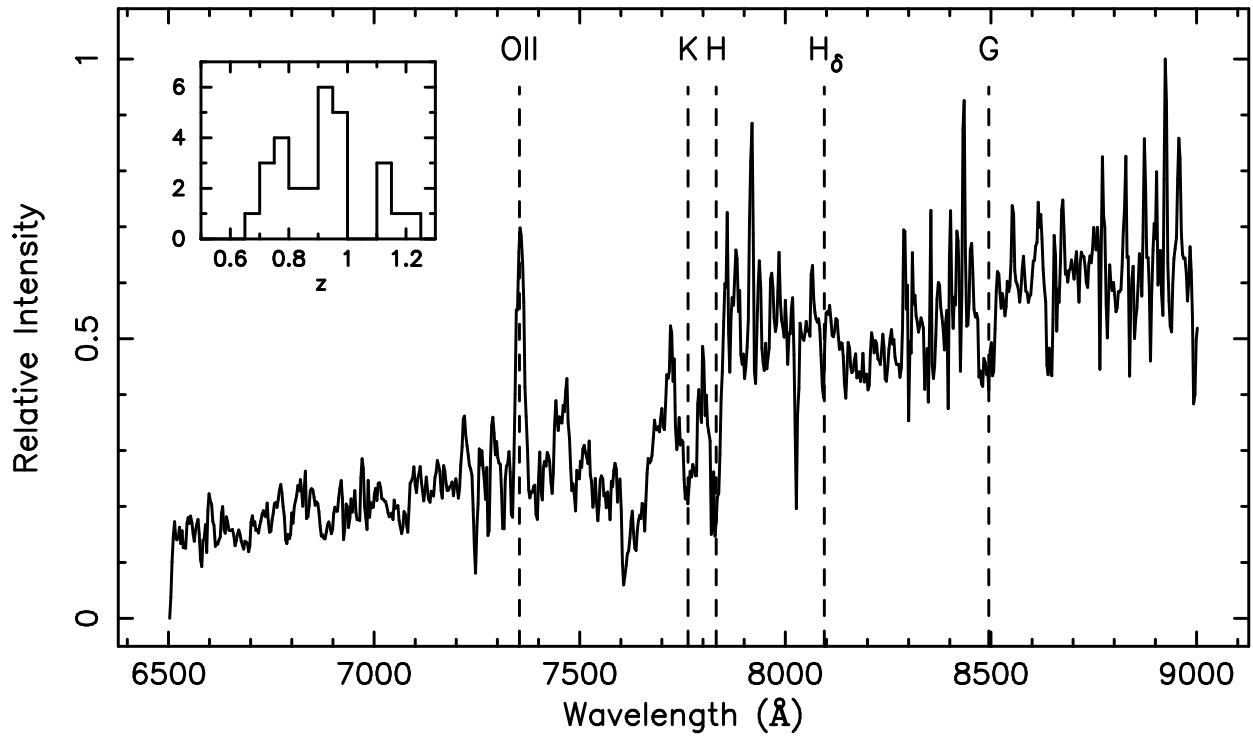}
\caption{VLT/FORS2 spectrum of one of the central galaxies in
RCS043938-2904.9, at $z=0.951 \pm 0.006$. The galaxy is marked in
Fig. 1 and has a redshift of $z=0.970$. Main features are indicated
with broken lines. The scale on the vertical axis is in arbitrary
units. The inset shows the redshift distribution in the field of this
cluster.}
\end{figure}

\begin{figure}
\plotone{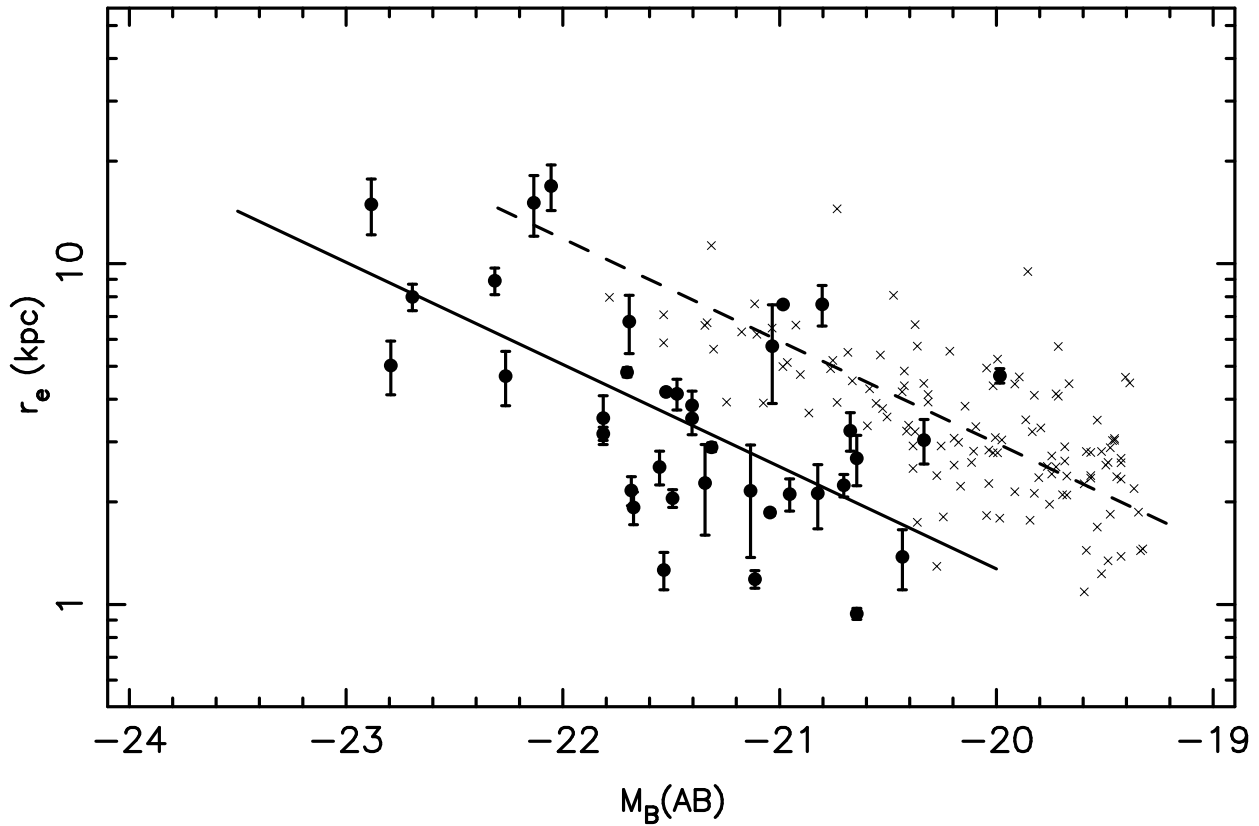}
\caption{Size-magnitude diagram for the E/S0 galaxies (filled circles)
in the field of \cluster. The size has been obtained from the 2-D
galaxy light profile fitting algorithm. For comparison the E/S0
galaxies in local clusters are shown as crosses, including the best
linear fit to these galaxies \citep[broken line,][]{schade1997b}. The
solid line corresponds to the fit for the galaxies in \cluster,
constrained to have the same slope as that for the local E/S0
galaxies. The offset of the fit for \cluster\ from the local relation
amounts to $\Delta M_B (AB) = -1.20 \pm 0.09$.}
\end{figure}

\begin{figure}
\plotone{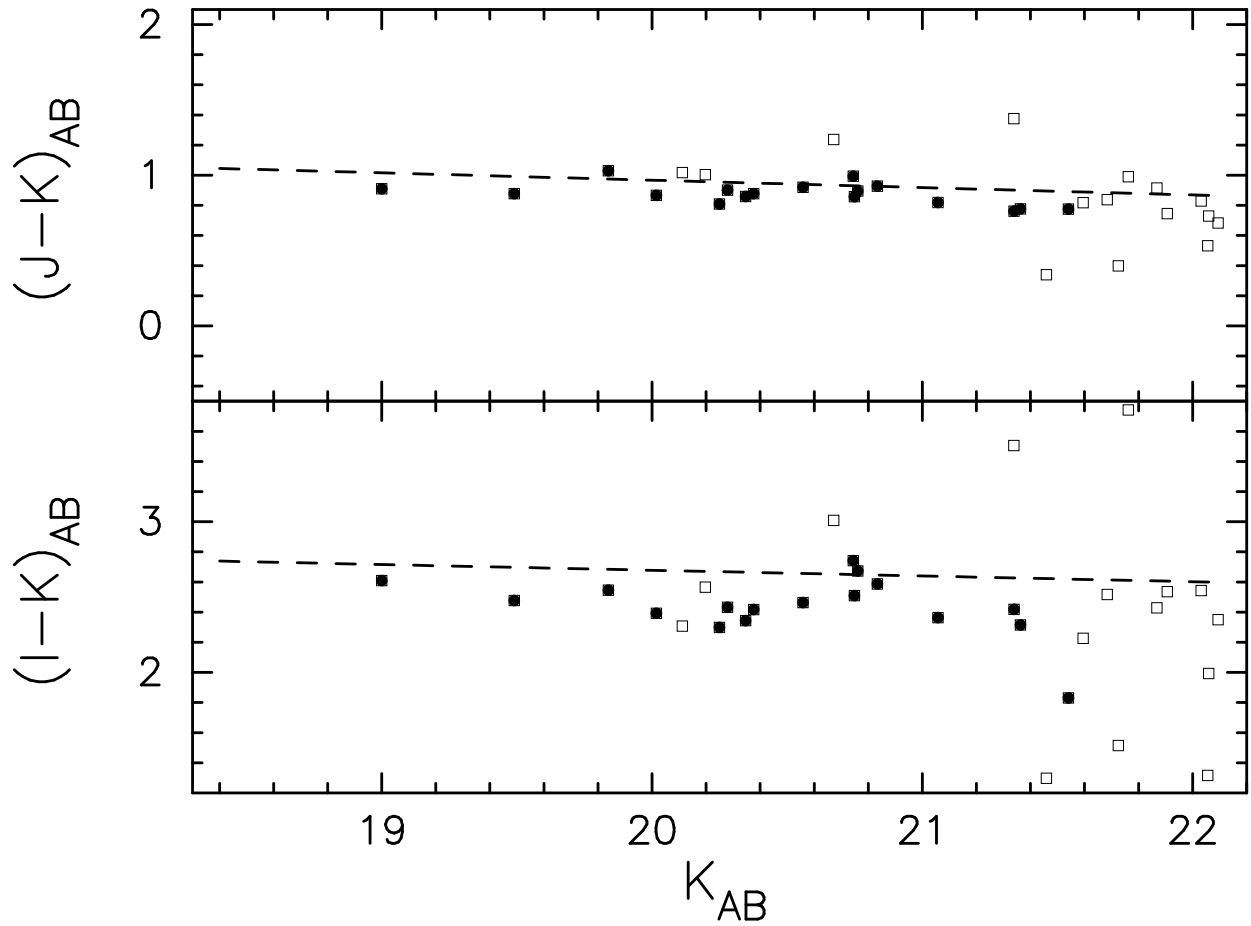}
\caption{IR and IR-optical color-magnitude diagrams in a 0.5 Mpc
diameter field centered on the cluster \cluster. The morphologically
selected E/S0 galaxies are shown as filled symbols. These galaxies
define a tight sequence (particularly in the IR), similar to that
found in local clusters. In this Figure, the expected color-magnitude
relation for the E/S0 galaxies in Coma as determined by
\citet{bower1992} and $k$-corrected to \clusterz, is shown as a broken
line.}
\end{figure}

\end{document}